# Broadband spintronic detection of the absolute field strength of terahertz electromagnetic pulses

A.L. Chekhov, Y. Behovits, U. Martens, B.R. Serrano, M. Wolf,
T.S. Seifert, M. Muenzenberg, and T. Kampfrath

**Abstract**
We demonstrate detection of broadband intense terahertz electromagnetic pulses by Zeeman-torque sampling (ZTS). Our approach is based on magneto-optic probing of the Zeeman torque the terahertz magnetic field exerts on the magnetization of a ferromagnet. Using an 8 nm thick iron film as sensor, we detect pulses from a silicon-based spintronic terahertz emitter with bandwidth 0.1-11 THz and peak field >0.1 MV/cm. Static calibration provides access to absolute transient THz field strengths. We show relevant added values of ZTS compared to electro-optic sampling (EOS): an absolute and echo-free transfer function with simple frequency dependence, linearity even at high terahertz field amplitudes, the straightforward calibration of EOS response functions and the modulation of the polarization-sensitive direction by an external AC magnetic field. Consequently, ZTS has interesting applications even beyond the accurate characterization of broadband high-field terahertz pulses for nonlinear terahertz spectroscopy.

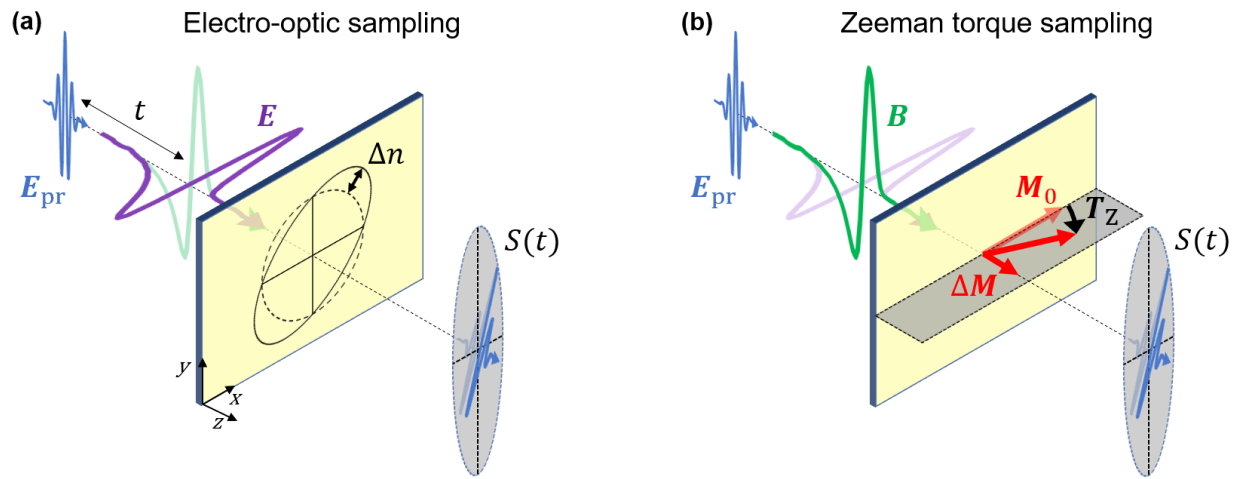

Figure 1. Schematics of THz electromagnetic-field detection by (a) electro-optic sampling (EOS) and (b) Zeeman-torque sampling (ZTS). In EOS, the THz electric field $E$ induces an anisotropic change in the refractive index $\Delta n$ and, thus, a linear optical birefringence of the detection material (dashed circle and solid ellipse). In contrast, in ZTS, the THz magnetic field $B$ exerts Zeeman torque $T_Z$ on the magnetization $M$ of the magnetic sample and, thus, induces a transient circular optical birefringence (Faraday effect).

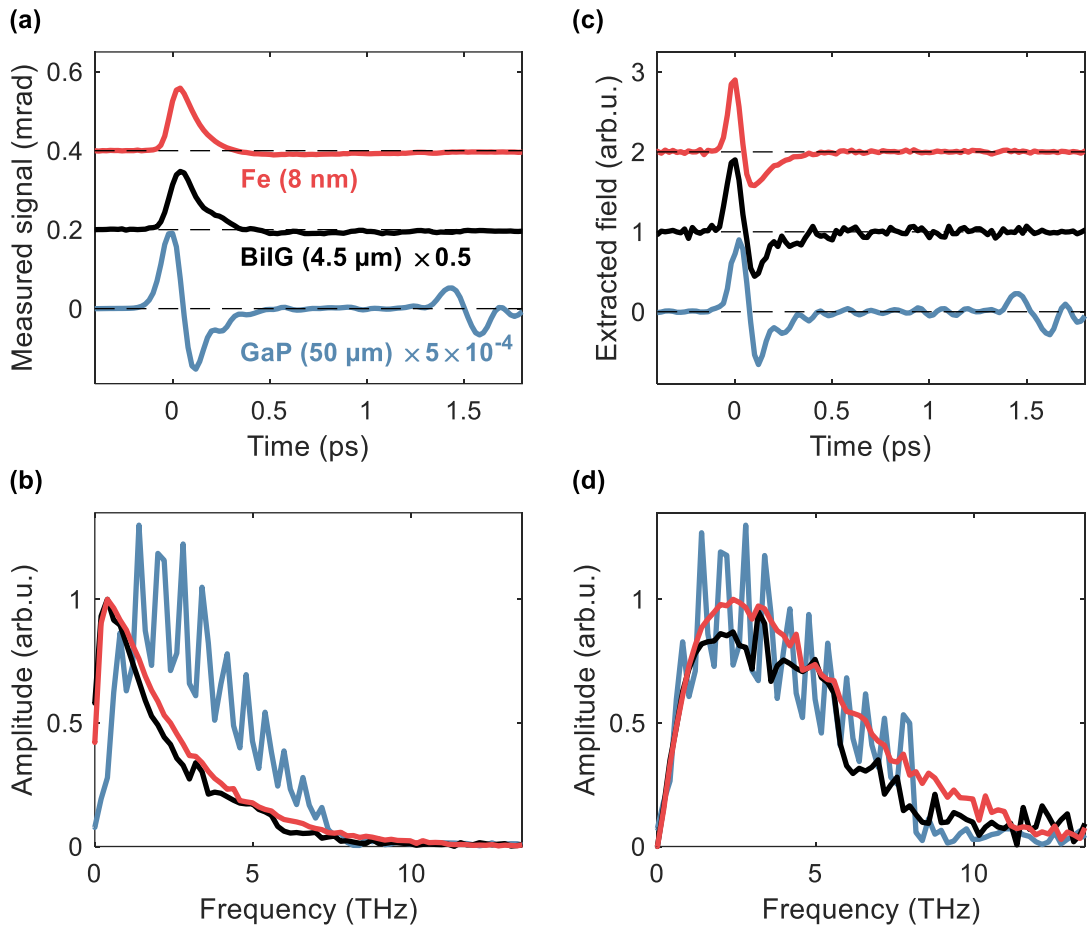

Figure 2. ZTS and EOS data. (a) Birefringence signals induced by the THz pump pulse obtained via ZTS in Fe(8 nm) (red solid line) and BiIG(4.5 μm) (black) and, for reference, by EOS in GaP(50 μm) (blue). (b) Fourier amplitude spectra of the waveforms of panel (a). (c) Transient THz electric fields, obtained by taking the time derivative of the ZTS signals (Fe and BiIG detectors) and by using the calculated EOS transfer function (reference GaP detector). Signals are normalized to their peak value. (d) Fourier amplitude spectra of the waveforms of panel (c).

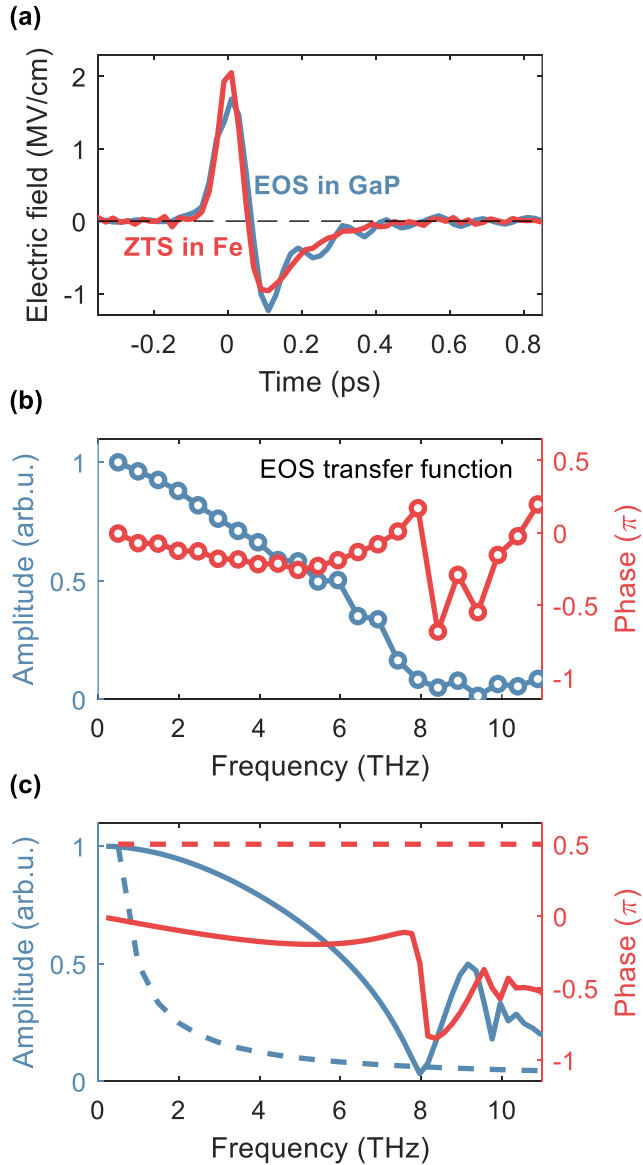

Figure 3. EOS characterization by ZTS. (a) Back-to-back comparison of the transient incident THz electric field obtained from ZTS and EOS signals of Fe(8 nm) and GaP(50 μm) in Fig. 2. (b) Amplitude (red) and phase (blue) of the EOS transfer function $\widetilde{H}_{\mathrm{GaP}}(\omega)$ for GaP(50 μm) obtained by referencing to the ZTS signal from Fe(8 nm). (c) Calculated response functions $\widetilde{H}_{\mathrm{GaP}}(\omega)$ (solid curves) and $\widetilde{H}_{\mathrm{Fe}}(\omega)$ (dashed curves).

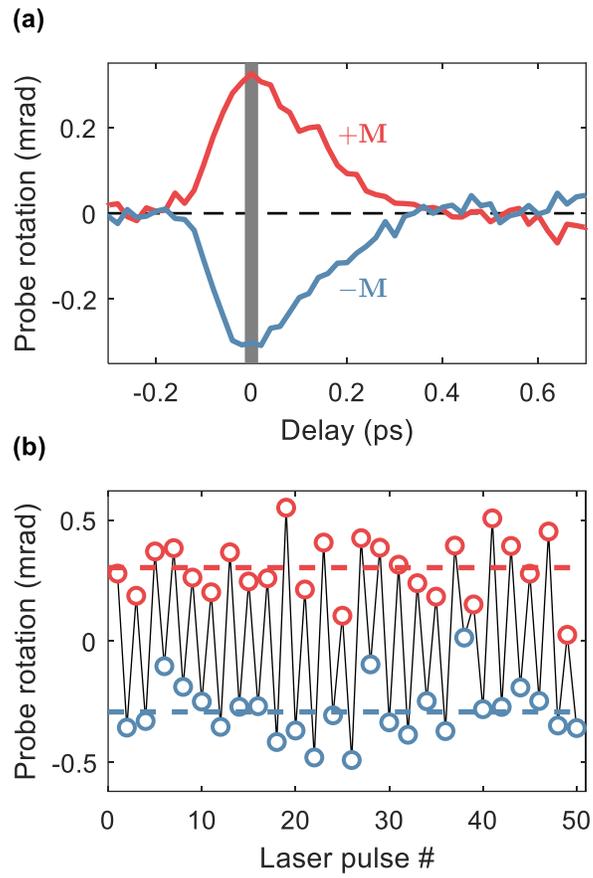

Figure 4. Rapid amplitude modulation of the ZT detector. (a) ZTS signal of a BiIG sample for opposite magnetization directions. (b) Pulse-to-pulse probe rotation, when the garnet magnetization is modulated at half the laser repetition rate of $f_{rep}/2 = 500$ Hz. The delay line is set to the maximum of the ZTS signal [see grey mark in panel (a)].

## Introduction

Recently developed high-field table-top terahertz (THz) sources [1–6] allow one to resonantly drive numerous low-energy excitations in all phases matters into previously inaccessible regimes [7–9]. Examples include the nonlinear excitation of magnons [10], nonlinear magnon dynamics [11–13] magnon-phonon coupling [14], the nonlinear THz Kerr effect in liquids [15] and the phononic switching [16,17] or generation [18] of magnetic order. For such experiments, it is essential to accurately determine the time-dependent electric field $E$ and magnetic field $B$ of the driving THz pulse at a given position.

The state-of-the-art method for THz-pulse detection is electro-optic sampling (EOS). It is based on the linear electro-optic (EO) effect in the sensing material, i.e., an anisotropic variation of the material's refractive index along certain eigendirections (linear birefringence), which is proportional to the applied transient electric field. EOS requires crystals with broken inversion symmetry, such as ZnTe, GaP or GaAs. The THz-field-induced transient birefringence $S(t)$ is typically measured by an optical probe pulse as a function of the delay $t$ between THz and optical probe pulses [Fig. 1(a)].

Determination of $E(t)$ from $S(t)$ requires accurate knowledge of the EOS transfer function, which is strongly frequency-dependent in the vicinity of phonon resonances, typically in the range 5-15 THz [19]. As a result, the extraction of the transient THz electric field $E(t)$, in particular absolute field strengths, from the measured $S(t)$ is nontrivial. This issue is most severe when the THz pulses cover large frequency intervals of, e.g., 1-10 THz and beyond, like those obtained by optical rectification in organic crystals [20] or from photocurrents in spintronic THz emitters [6,21].

An interesting alternative route to THz-pulse detection is offered by the THz magnetic field $B(t)$. In ferro- or ferrimagnets, $B$ straightforwardly couples to the magnetization $M$ of the material by Zeeman torque (ZT) and, thus, drives ultrafast magnetization dynamics [22–28]. So far, ZT was used to manipulate $M$ with bandwidths up to 5 THz [25].

Here, we show that the magnetization dynamics in a ferro-/ferrimagnet driven by the ZT of the magnetic component of THz pulses can be used to measure intense transient THz electric fields of $> 0.1$ MV/cm over the entire range 0.1-11 THz. We optimize the THz ZT detector composition to achieve the best performance in terms of signal-to-noise ratio and bandwidth. The response function of ZT sampling (ZTS) is found to have a simple frequency dependence, and its absolute magnitude can be calibrated by a static magnetic field. As a first application, we use our detection method to determine the response function of commercial EOS crystals. Second, we demonstrate modulation of the Zeeman-detector response at kilohertz rates with an external magnetic field, thereby enabling demodulation schemes of THz waves.

## ZT detection

In our ZTS approach [Fig. 1(b)], the THz pulse to be measured and an optical probe pulse are normally incident on a magnetic thin film, the static magnetization $M_0$ of which lies in the film plane. The magnetic field $B(t)$ of the THz pulse exerts a torque proportional to $M \times B$ and, thus, deflects the dynamic magnetization $M$ out of the plane. For sufficiently small fields, the magnetization change $\Delta M = M - M_0$ is given by

$$\Delta M(t) = \gamma M_0 \times \int_{-\infty}^{t} \mathrm{d}t' \, B(t'), \tag{1}$$

where $\gamma$ is the electron gyromagnetic ratio. The optical probe pulse detects the out-of-plane component of $\Delta M$, averaged over the probed volume, by the Faraday effect, i.e., the circular birefringence induced by $\Delta M$. As shown in Appendix A, the birefringence signal at THz frequency $\omega/2\pi$ is given by

$$\tilde{S}(\omega) = \tilde{H}(\omega)\boldsymbol{v} \cdot \tilde{\boldsymbol{E}}_{\text{inc}}(\omega). \tag{2}$$

It is proportional to the projection of the incident THz field on the unit-vector direction $\boldsymbol{v} = \boldsymbol{M}_0/|\boldsymbol{M}_0|$ of the static magnetization. The transfer function

$$\tilde{H}(\omega) = C_E(\omega)C_{\text{pr}} \tag{3}$$

has two contributions. First, as detailed in Appendix A, the coefficient $C_E(\omega) \propto 1/i\omega$ [Eq. (13)] quantifies the coupling of the incident THz electromagnetic field into the thin film and the coupling strength of $\boldsymbol{B}$ to $\boldsymbol{M}$, where the straightforward $1/i\omega$ frequency dependence arises from the time integration of $\boldsymbol{B}$ and, thus, $\boldsymbol{E}$ [see Eq. (1)]. Second, the coefficient $C_{\text{pr}}$ is independent of $\omega$ [Eq. (13)] and describes the probe coupling to $\Delta\boldsymbol{M}$ and coupling out of the film.

Note that Eq. (3) is an approximation (see Appendix A), which is valid for metallic Zeeman detectors. Interestingly, in this approximation, the ZT signal $S(t)$ is proportional to the time integral $\int_{-\infty}^{t} dt' \, \boldsymbol{E}_{\text{inc}}(t')$ and, thus, the vector field of the incident THz pulse in radiation gauge, similar to signals obtained by attosecond sampling of light fields [29].

We emphasize that the magnitude of the transfer function $\tilde{H}(\omega)$ can be determined by a straightforward calibration measurement (Appendix B), thereby providing access to the *absolute* strength of the incident THz electric field.

**Experimental details**

**THz setup.** To test the capabilities of ZTS [Fig. 1(a)], we use intense broadband THz pulses (spectrum 0.1-11 THz, peak field 1.7 MV/cm) from a Si-based spintronic THz emitter (Si-STE, TeraSpinTec GmbH) [6] and sample their focal field with probe pulses (center wavelength 800 nm, duration 30 fs) in the ZT detector (see Fig. S1 for details).

For referencing, we use EOS in a freestanding GaP(110) crystal (Swiss THz), whose thickness of only 50 μm facilitates a wide detection window of 0-7 THz and beyond [19]. The EOS response function is analogous to Eq. (2). The transfer function $\tilde{H}(\omega)$ and the unit-length vector $\boldsymbol{v}$ are just replaced by expressions that are determined by the refractive index and quadratic optic response tensor of GaP as well as the GaP thickness, azimuthal orientation, the probe pulse's electric field and its linear polarization state [6,30,31].

**ZTS detectors.** Because $C_{\text{pr}}$ is proportional to the thickness $d$ and the magneto-optic Voigt parameter $Q \propto |\boldsymbol{M}_0|$ of the ZT film, the ZTS response is enhanced for films with large $Q$ and optimum thickness. The latter also depends on the magnitude of the probe transmission. Consequently, we choose the metallic ferromagnet Fe and the insulating ferrimagnetic bismuth-substituted iron garnet (BiIG).

Fe features the largest Voigt parameter $|Q| \sim 10^{-2}$ among 3d ferromagnets around the 800 nm probe wavelength [32], but, at the same time, strongly attenuates the probe beam [33]. We, therefore, choose an optimum thickness $d$ of 8 nm. In contrast, BiIG has a 1 order of magnitude smaller $|Q| \sim 10^{-3}$, but 3 orders of magnitude smaller absorption coefficient and less reflection losses than Fe [34]. At the same time, phonon resonances in the garnet can distort and attenuate the THz pulse [35]. To minimize this effect, we choose a thickness of 4.5 μm.

The Fe(8 nm) film is obtained by electron-beam evaporation under ultrahigh-vacuum conditions onto a glass substrate. The BiIG(4.5 μm) film (INNOVENT e.V.) with in-plane magnetic anisotropy is grown on a gallium gadolinium garnet substrate by liquid-phase epitaxy. The magnetization $\boldsymbol{M}_0$ is set by a Halbach array of permanent magnets and can be reversed by 180° azimuthal rotation [6].

**Calibration.** To calibrate our detector, the most straightforward approach would be to completely pull the static magnetization of the ferromagnet out of the plane and measure the resulting Faraday rotation and ellipticity of the probe beam. For our in-plane magnetized Fe(8 nm) film, this procedure requires static magnetic fields $> 2$ T, which are not easily available.

Therefore, we apply an in-plane magnetic field along the $x$-axis [Fig. 1(b)] and measure the magneto-optic response for an s-polarized probe beam in a reflection geometry (angle of incidence 52°), i.e., the so-called longitudinal magneto-optic Kerr effect (LMOKE). As detailed in Appendix B, this information, along with the optical and THz refractive indices (or conductivity) of Fe and glass (see Table S1), is sufficient to determine the absolute magnitude of $\widetilde{H}(\omega) \propto 1/i\omega$.

**Signal contributions.** In the experiment, the intense THz pulses can induce contributions nonlinear in the THz fields [24,36]. To discriminate quadratic contributions, we measure pump-probe signals $S_{\pm E,\pm M_0}$ for opposite polarities of the THz field ($\pm E$) by reversing the Si-STE magnetization direction. Furthermore, to separate effects odd (linear) in the ZT detector magnetization $M_0$, we alternate it ($\pm M_0$) by means of an AC electromagnet at a frequency of 500 Hz, i.e., half the repetition rate of the THz pulses. The extracted signals $S = S(t)$ linear in both the driving THz field $E$ and the ZT magnetization $M_0$ are calculated as $S = (S_{+E,+M_0} - S_{-E,+M_0} - S_{+E,-M_0} + S_{-E,-M_0})/4$ (see Fig. S2).

### ZTS raw data

Fig. 2(a) shows signal waveforms $S(t)$ obtained with the various ZT and EO detectors under identical THz-pulse excitation. While the EO signal from GaP and the ZT signal from Fe correspond to probe ellipticity variation, the ZT signal in BiIG was measured as probe polarization rotation (see Fig. S3). A Fourier transformation yields the corresponding signal amplitude spectra shown in Fig. 2(b). For the Fe sample, we confirmed that the measured signal $S(t)$, which is odd in $E$, grows linearly with $E$. Contributions quadratic in $E$ are minor, in particular ultrafast quenching of the in-plane magnetization, which is not sensed by the Faraday effect of the normally incident probe pulses (see Fig. S2).

The EO signal from the GaP(50 µm) reference detector exhibits a main pulse centered at time $t = 0$ ps, but also a sizeable echo at $t = 1.5$ ps [Fig. 2(a)], which results in spectral oscillation with period $1/1.5$ ps $= 0.7$ THz [Fig. 2(b)]. This echo arises from reflection of the THz pulse at the GaP/air and air/GaP interfaces and limits the time window and, thus, frequency resolution of EOS.

Note that the ZT signals are substantially different from the EO reference signal. First, no echo is found in the covered time window. Second, the ZT signals are reminiscent of the time-integrated EO signal [Fig. 2(a)], as expected from Eq. (1). Interestingly, the ZT signal from Fe is only a factor of 2 smaller than from BiIG, but has a somewhat smoother time [Fig. 2(a)] and frequency dependence [Fig. 2(b)].

### From signals to incident fields

To extract the THz-field transients from the measured EO and ZT signals of Fig. 2(a), we employ Eq. (2). For EOS, $\widetilde{H}(\omega)$ is obtained from previous work [21,22], while for ZTS, we only need to take the time derivative $\partial_t S(t)$ of the measured signal according to Eq. (3).

The resulting normalized THz electric-field transients are shown in Fig. 2(c). They all exhibit a very similar shape, indicating a reliable extraction procedure. As expected from the ZT signals [Fig. 2(a)], the field trace obtained with Fe(8 nm) is significantly smoother than the trace obtained with BiIG(4.5 µm). We ascribe this behavior to two effects: (i) a substantial variation of the BiIG refractive index $n(\omega)$ in the considered frequency interval 0-12 THz due to the large number of infrared-active optical phonon modes [35], and

(ii) long-wavelength quasi-antiferromagnetic magnons [38] that may alter the simple $1/i\omega$ scaling of $C_E$ in Eq. (3). Points (i) and (ii) could be accounted for by an appropriate frequency dependence of $C_E$.

We note that the refractive index of Fe exhibits dispersion, too. However, as shown in Appendix A and Fig. S6, the resulting frequency dependence of $C_E(\omega)$ in Eq. (3) can still be approximated by $1/i\omega$. To enable ZT detection of broadband THz pulses with a smooth frequency response, we focus on Fe in the remainder of this work.

**Extracting the absolute THz field**

As detailed in Appendix B, we use static LMOKE measurement to calibrate the Fe(8 nm) ZTS detector. To very good approximation, we find $\widetilde{H}(\omega) = b/i\omega$ with $b = -6.9 + 11.1i$ m/V.

Subsequently, we use Eq. (2) to extract the incident THz electric field $E_{\mathrm{inc}}(t) = \boldsymbol{v} \cdot \boldsymbol{E}_{\mathrm{inc}}(t)$ with $\boldsymbol{v} \parallel \boldsymbol{E}_{\mathrm{inc}}$ from the measured probe ellipticity variation. The resulting $E_{\mathrm{inc}}(t)$ is in good agreement with the field extracted from EOS in the GaP detector, both in terms of shape and absolute amplitude [Fig. 3(a)].

**ZTS applications**

**THz field sampling.** Our ZT THz detector Fe(8 nm) permits accurate sampling of intense THz fields (strength > 0.1 MV/cm) over a bandwidth 0.1-11 THz and beyond, without showing indications of saturation (see Fig. S2). This feature is crucial for virtually all nonlinear THz-spectroscopy experiments.

As a further proof of principle, we use our ZT detector for sampling of THz pulses generated through tilted-pulse-front-excitation of LiNbO$_3$ (Fig. S4). Again, the extracted THz electric field is in good agreement with the one obtained by EOS.

**EOS calibration.** As a second application, we use our ZT detector to calibrate THz EO detectors. So far, the response function $\widetilde{H}(\omega)$ of EO detectors with zincblende structure, e.g., GaP and ZnTe, are calculated based on models with a number of parameters such as ionic vs electronic nonlinearity, whose determination is not straightforward [19,39].

To determine $\widetilde{H}(\omega)$ of, for example, GaP(50 μm), we take the THz signals measured by ZTS in Fe(8 nm) and by EOS in GaP(50 μm), make use of Eq. (2) and the fact that the experimental THz excitation and optical probing conditions were the same. We obtain the EO transfer function

$$\widetilde{H}_{\mathrm{GaP}}(\omega) = \frac{\widetilde{S}_{\mathrm{GaP}}(\omega)}{\widetilde{S}_{\mathrm{Fe}}(\omega)} \widetilde{H}_{\mathrm{Fe}}(\omega), \tag{4}$$

the frequency dependence of which directly follows from the measured signals $S_{\mathrm{GaP}}$ and $S_{\mathrm{Fe}}$ and Eq. (3). The spectral amplitude and phase of the resulting $\widetilde{H}_{\mathrm{GaP}}(\omega)$ [Fig. 3(b)] agree well with model calculations [Fig. 3(c)]. For comparison, the calculated $\widetilde{H}_{\mathrm{Fe}}(\omega)$ is shown in Fig. 3(c), too.

**ZTS modulation.** A third outstanding functionality of the ZT detector follows from Eq. (2). The measured signal is proportional to the projection $\boldsymbol{v} \cdot \boldsymbol{E}_{\mathrm{inc}}$ of the incident THz electric field $\boldsymbol{E}_{\mathrm{inc}}$ on the magnetization direction $\boldsymbol{v} = \boldsymbol{M}/|\boldsymbol{M}_0|$. Therefore, we can easily modulate the sign of $\widetilde{H}_{\mathrm{Fe}}(\omega)$ by modulating the magnetization of the ZT detector between $-\boldsymbol{M}_0$ and $+\boldsymbol{M}_0$.

To demonstrate this THz demodulation scheme, we employ an AC electromagnet operating at a frequency $f_{\mathrm{rep}}/2 = 500$ Hz, i.e., half the repetition rate of the THz pulses [36]. The delay time $t$ is set to the maximum of the ZT signal for one magnetization direction [Fig. 4(a)]. Thus, we modulate the Fe magnetization between $\pm \boldsymbol{M}_0$ at $f_{\mathrm{rep}}/2$, and signals are measured for each THz pulse at the rate $f_{\mathrm{rep}}$. Fig. 4(b) displays the signal recorded for some 50 subsequent pulses. The alternating signal amplitude clearly confirms the amplitude modulation of the ZT detector response at $f_{\text{rep}}/2 = 500$ Hz. While this frequency is limited by the laser-pulse repetition rate, modulation frequencies of several 10 kHz or even 1 MHz are feasible by sufficiently fast magnets [40], strain waves [41] or current-driven spin-orbit torques [42].

**Conclusions**

In conclusion, we demonstrate the detection of transient THz electric fields by ZTS. The ZT detector has outstanding features that enable interesting applications, for example, sampling of intense THz pulses with strength $> 0.1$ MV/cm with straightforward gapless transfer function and known absolute magnitude, characterization of the response function of EO detectors and the modulation of the polarization sensitivity in a non-contact fashion by an external magnetic field.

The last feature is interesting for demodulation of modulated THz waves in, for instance, low-noise lock-in detection schemes or data transfer. Our study may serve as a guide of future works that aim to increase the sensitivity of ZT detectors.

## Appendix A. Derivation of ZTS response function for thin metallic films.

The magnetic field $\boldsymbol{B}$ of the THz pulse inside the magnetic film exerts Zeeman torque on the magnetization $\boldsymbol{M}$. We assume small magnetization deflections [$\boldsymbol{M} \approx \boldsymbol{M}_0$ on the right-hand side of Eq. (1)] and only consider free uniform precession. In particular, we neglect effects such as perpendicular standing spin waves, attenuation and nutation [28,43,44]. In the frequency domain, the resulting out-of-plane magnetization component is given by the Fourier transformation of Eq. (1),

$$\Delta \widetilde{\boldsymbol{M}}(\omega, z) = \frac{\gamma}{\mathrm{i}\omega} \boldsymbol{M}_0 \times \widetilde{\boldsymbol{B}}(\omega, z). \tag{5}$$

Assuming the magnetic medium has cubic symmetry for $\boldsymbol{M}_0 = 0$, the transient out-of-plane magnetization gives rise to off-diagonal elements $\Delta \tilde{\varepsilon}_{yx}(\omega, z) = -\Delta \tilde{\varepsilon}_{xy}(\omega, z)$ of the dielectric permittivity tensor, where $\Delta \tilde{\varepsilon}_{yx}(\omega, z) = \mathrm{i} Q \varepsilon_2 \boldsymbol{u}_z \cdot \Delta \widetilde{\boldsymbol{M}}(\omega, z)/|\boldsymbol{M}_0|$, $\boldsymbol{u}_z$ is the unit vector along the $z$-axis, $Q \propto |\boldsymbol{M}_0|$ is the Voigt magneto-optic coefficient, and $\varepsilon_2$ is the dielectric permittivity of the metal film at the mean probe frequency. The element $\Delta \tilde{\varepsilon}_{yx}(\omega, z)$ results in magnetic circular birefringence (Faraday effect) and modulates the probe polarization.

To calculate the probe polarization change, we consider $\Delta \tilde{\varepsilon}_{yx}(\omega, z)$ as a small perturbation [45,46]. As the film is thin with respect to the THz and visible wavelengths and because the probe resides within the film for a minor amount of time, we assume instantaneous sampling of the entire film by the probe. Likewise, we neglect the dependence of the optical coefficients on the probe frequency $\omega_{\mathrm{pr}}$ over the probe spectrum. With the unperturbed probe field with amplitude $\boldsymbol{E}_{\mathrm{pr0}}(\omega_{\mathrm{pr}}, z)$ linearly polarized along the $x$-axis, we can obtain the first-order change $\Delta \boldsymbol{E}_{\mathrm{pr}}(\omega_{\mathrm{pr}}, z)$ in the probe field polarized along the $y$-axis as [45]

$$\Delta \tilde{E}_{\mathrm{pr}}(\omega_{\mathrm{pr}}, z) = -\frac{\omega_{\mathrm{pr}}^2}{c^2} \int_0^d \mathrm{d}z' \, G_{z'}(\omega_{\mathrm{pr}}, z) \Delta \tilde{\varepsilon}_{yx}(\omega, z') E_{\mathrm{pr0}}(\omega_{\mathrm{pr}}, z'). \tag{6}$$

Here, $d$ is the thickness of the metal film, $G_{z'}(\omega_{\mathrm{pr}}, z)$ is the known electric-field Green's function of the structure for $\tilde{\varepsilon}_{yx} = 0$. After substituting the corresponding expressions for $G_{z'}(\omega_{\mathrm{pr}}, z)$ and $E_{\mathrm{pr0}}(\omega_{\mathrm{pr}}, z')$ [46], we obtain the complex-valued probe-polarization change behind the magnetic film ($z = d$) according to

$$\tilde{S}(\omega) = \frac{\Delta \tilde{E}_{\mathrm{pr}}(\omega_{\mathrm{pr}}, d)}{\tilde{E}_{\mathrm{pr0}}(\omega_{\mathrm{pr}}, d)} = \frac{\mathrm{i}\omega_{\mathrm{pr}}^2}{2\beta_2 c^2} R(\omega_{\mathrm{pr}}) \int_0^d \mathrm{d}z \, \Delta \tilde{\varepsilon}_{yx}(\omega, z) \left(1 + r_{21} \mathrm{e}^{2\mathrm{i}\beta_2 z}\right)\left(1 + r_{23} \mathrm{e}^{2\mathrm{i}\beta_2 (d-z)}\right), \tag{7}$$

where $\beta_i(\omega_{\mathrm{pr}}) = n_i(\omega_{\mathrm{pr}}) \omega_{\mathrm{pr}}/c$ is the angular spatial frequency of medium $i$ (air, metal, substrate), $r_{ij} = (n_i - n_j)/(n_i + n_j)$ are Fresnel reflection coefficients, and $R = 1/(1 - r_{21} r_{23} \mathrm{e}^{2\mathrm{i}\beta_2 d})$ accounts for the multiple probe reflections in the metal film. As shown in Fig. S5, the last 2 factors of the integrand in Eq. (7) can be approximated well by their values at $z = d/2$ for the parameter values shown in Table 1. Considering Eq. (5) and expressing $\varepsilon_{yx}$ through the Voigt parameter, we arrive at

$$\tilde{S}(\omega) = -\frac{Q \beta_2}{2} \frac{\left(1 + r_{21} \mathrm{e}^{\mathrm{i}\beta_2 d}\right)\left(1 + r_{23} \mathrm{e}^{\mathrm{i}\beta_2 d}\right)}{1 - r_{21} r_{23} \mathrm{e}^{2\mathrm{i}\beta_2 d}} \frac{\gamma}{\mathrm{i}\omega} \int_0^d \mathrm{d}z \, \boldsymbol{u}_z \cdot \left(\boldsymbol{v} \times \widetilde{\boldsymbol{B}}(\omega, z)\right) \tag{8}$$

with $\boldsymbol{v} = \boldsymbol{M}_0/|\boldsymbol{M}_0|$.

To determine $\widetilde{\boldsymbol{B}}(\omega, z)$, we note that the studied Fe film has a thickness $d$ much smaller than the THz wavelength and the THz attenuation length inside the film material. Therefore, the THz electric field is to a very good approximation constant across the film thickness. Its amplitude at frequency $\omega$ is given by

$$\widetilde{E}(\omega) = \frac{2n_1(\omega)}{n_1(\omega) + n_3(\omega) + dZ_0\sigma(\omega)} \widetilde{E}_{\text{inc}}(\omega), \tag{9}$$

where $\sigma(\omega)$ is the conductivity of the metal film, and $Z_0$ is the vacuum impedance. By using Eq. (9) and Ampere's law, we obtain the magnetic field

$$\widetilde{\boldsymbol{B}}(\omega, z) = \frac{2}{c} \frac{n_3(\omega) + (d-z)Z_0\sigma(\omega)}{n_1(\omega) + n_3(\omega) + dZ_0\sigma(\omega)} \boldsymbol{u}_z \times \widetilde{\boldsymbol{E}}_{\text{inc}}(\omega), \tag{10}$$

and Eqs. (8) and (10) yield

$$\widetilde{S}(\omega) = -\frac{Q\beta_2 d}{2} \frac{(1+r_{21}e^{i\beta_2 d})(1+r_{23}e^{i\beta_2 d})}{1-r_{21}r_{23}e^{2i\beta_2 d}} \frac{\gamma}{i\omega c} \frac{2n_3(\omega) + Z_0 d\sigma(\omega)}{n_1(\omega) + n_3(\omega) + Z_0 d\sigma(\omega)} \boldsymbol{v} \cdot \widetilde{\boldsymbol{E}}_{\text{inc}}(\omega). \tag{11}$$

We note that the term in front of $\boldsymbol{v} \cdot \widetilde{\boldsymbol{E}}_{\text{inc}}$ in Eq. (11) is constant over the whole bandwidth 0.1-10 THz (Fig. S6) and can be well approximated by its value at $\omega = 0$ for the parameters listed in Table S1.

As a consequence, the ZTS response finally becomes

$$\widetilde{S}(\omega) = C_{\text{pr}} C_E(\omega) \boldsymbol{v} \cdot \widetilde{\boldsymbol{E}}_{\text{inc}}(\omega) \tag{12}$$

with

$$C_{\text{pr}} = -\frac{Q\beta_2 d}{2} \frac{(1+r_{21}e^{i\beta_2 d})(1+r_{23}e^{i\beta_2 d})}{1-r_{21}r_{23}e^{2i\beta_2 d}}, \quad C_E(\omega) = \frac{\gamma}{i\omega c} \frac{2n_3(\omega) + Z_0 d\sigma_0}{n_1(\omega) + n_3(\omega) + Z_0 d\sigma_0}. \tag{13}$$

Note that the prefactor $-Q\beta_2 d/2$ is the complex-valued Faraday rotation for a thick magnetic film [32].

**Appendix B. Zeeman detector calibration.**

As seen from Eqs. (12) and (13), the transfer function of a metal-film-based Zeeman detector is fully determined by its THz and optical refractive indices, THz conductivity and magneto-optic Voigt parameter. In contrast to EOS crystals, all required values can be measured using linear optical and THz spectroscopies. Here, we limit ourselves to the experimental determination of the magneto-optic Voigt parameter $Q$, while all other values are taken from the literature (see Table S1) [36,47,48].

In the experiment, we use a reflection geometry (angle of incidence 52°, s-polarized probe) and magnetize our Fe sample along the $x$-axis [Fig. 1(b)], corresponding to the so-called longitudinal magneto-optic Kerr-effect (LMOKE) geometry. By measuring both rotation and ellipticity variations of the probe polarization, we obtain the complex-valued LMOKE signal $S = 0.0016 - 0.0011i$ (see Fig. S7). As the magneto-optic Voigt parameter $Q$ is small, the response is linear, i.e., $S = aQ$ with coefficient $a$. Using a $4 \times 4$-transfer-matrix formalism [49,50], we numerically determine the value of $a$ for a Fe film on a glass substrate using the parameter values listed in Table S1. Finally, we obtain $Q = S/a = 0.060 - 0.001i$, calculate the ZTS transfer function $\widetilde{H}(\omega)$ [Eqs. (12) and (13)] and arrive at the result shown in Fig. 3(a). To very good approximation, we find $\widetilde{H}(\omega) = b/i\omega$ with $b = -6.1 + 11.1i$ m/V for our Fe(8 nm) ZTS detector.

It is important to note that both $Q$ and $n_2(\omega_{\text{pr}})$ were reported to be thickness-dependent in thin iron films [47,51]. Here, we use an effective refractive index value obtained for a Fe thin film with a thickness of 12 nm [47]. For even better results, it is advisable to determine the refractive index for the studied film separately. Additionally, one can use epitaxially grown films [51], where the impact of the thickness dependence is smaller compared to sputter-deposited samples [47].

**Acknowledgments**

The authors acknowledge funding by the DFG collaborative research center SFB TRR 227 "Ultrafast spin dynamics" (project ID 328545488, projects A05, B02 and B05), the DFG priority program SPP2314 INTEREST (project ITISA) and financial support from the Horizon 2020 Framework Program of the European Commission under FET Open Grant No. 863155 (s-Nebula).

T.S.S. and T.K. are shareholders of TeraSpinTec GmbH, and T.S.S. is an employee of TeraSpinTec GmbH.

**References**
[1]     J. Hebling , K.-L. Yeh, M. C. Hoffmann, B. Bartal, and K. A. Nelson, *Generation of High-Power Terahertz Pulses by Tilted-Pulse-Front Excitation and Their Application Possibilities*, JOSA B **25**, B6 (2008).
[2]     X. Ropagnol, M. Khorasaninejad, M. Raeiszadeh, S. Safavi-Naeini, M. Bouvier, C. Y. Côté, A. Laramée, M. Reid, M. A. Gauthier, and T. Ozaki, *Intense THz Pulses with Large Ponderomotive Potential Generated from Large Aperture Photoconductive Antennas*, Opt. Express **24**, 11299 (2016).
[3]     A. Nguyen, K. J. Kaltenecker, J.-C. Delagnes, B. Zhou, E. Cormier, N. Fedorov, R. Bouillaud, D. Descamps, I. Thiele, S. Skupin, P. U. Jepsen, and L. Bergé, *Wavelength Scaling of Terahertz Pulse Energies Delivered by Two-Color Air Plasmas*, Opt. Lett. **44**, 1488 (2019).
[4]     J. A. Fülöp, S. Tzortzakis, and T. Kampfrath, *Laser-Driven Strong-Field Terahertz Sources*, Adv. Opt. Mater. **8**, 1900681 (2020).
[5]     C. Rader, Z. B. Zaccardi, S. H. E. Ho, K. G. Harrell, P. K. Petersen, M. F. Nielson, H. Stephan, N. K. Green, D. J. H. Ludlow, M. J. Lutz, S. J. Smith, D. J. Michaelis, and J. A. Johnson, *A New Standard in High-Field Terahertz Generation: The Organic Nonlinear Optical Crystal PNPA*, ACS Photonics **9**, 3720 (2022).
[6]     R. Rouzegar, A. L. Chekhov, Y. Behovits, B. R. Serrano, M. A. Syskaki, C. H. Lambert, D. Engel, U. Martens, M. Münzenberg, M. Wolf, G. Jakob, M. Kläui, T. S. Seifert, and T. Kampfrath, *Broadband Spintronic Terahertz Source with Peak Electric Fields Exceeding 1.5 MV/cm*, Phys. Rev. Appl. **19**, 034018 (2023).
[7]     T. Kampfrath, K. Tanaka, and K. A. Nelson, *Resonant and Nonresonant Control over Matter and Light by Intense Terahertz Transients*, Nat. Photonics **7**, 680 (2013).
[8]     H. Y. Hwang, S. Fleischer, N. C. Brandt, B. G. Perkins, M. Liu, K. Fan, A. Sternbach, X. Zhang, R. D. Averitt, and K. A. Nelson, *A Review of Non-Linear Terahertz Spectroscopy with Ultrashort Tabletop-Laser Pulses* **62**, 1447 (2014).
[9]     D. Nicoletti and A. Cavalleri, *Nonlinear Light–Matter Interaction at Terahertz Frequencies*, Adv. Opt. Photonics **8**, 401 (2016).
[10]    S. Baierl, J. H. Mentink, M. Hohenleutner, L. Braun, T. M. Do, C. Lange, A. Sell, M. Fiebig, G. Woltersdorf, T. Kampfrath, and R. Huber, *Terahertz-Driven Nonlinear Spin Response of Antiferromagnetic Nickel Oxide*, Phys. Rev. Lett. **117**, 197201 (2016).
[11]    J. Lu, X. Li, H. Y. Hwang, B. K. Ofori-Okai, T. Kurihara, T. Suemoto, and K. A. Nelson, *Coherent Two-Dimensional Terahertz Magnetic Resonance Spectroscopy of Collective Spin Waves*, Phys. Rev. Lett. **118**, 207204 (2017).
[12]    Y. Behovits, A. L. Chekhov, S. Y. Bodnar, O. Gueckstock, S. Reimers, T. S. Seifert, M. Wolf, O. Gomonay, M. Kläui, M. Jourdan, and T. Kampfrath, arXiv:2305.03368.
[13]    Z. Zhang, F. Sekiguchi, T. Moriyama, S. C. Furuya, M. Sato, T. Satoh, Y. Mukai, K. Tanaka, T. Yamamoto, H. Kageyama, Y. Kanemitsu, and H. Hirori, *Generation of Third-Harmonic Spin Oscillation from Strong Spin Precession Induced by Terahertz Magnetic near Fields*, Nat. Commun. **14**, 1 (2023).
[14]    E. A. Mashkovich, K. A. Grishunin, R. M. Dubrovin, A. K. Zvezdin, R. V. Pisarev, and A. V. Kimel, *Terahertz Light–Driven Coupling of Antiferromagnetic Spins to Lattice*, Science **374**, 1608 (2021).
[15]    H. Elgabarty, T. Kampfrath, D. J. Bonthuis, V. Balos, N. K. Kaliannan, P. Loche, R. R. Netz, M. Wolf, T. D. Kühne, and M. Sajadi, *Energy Transfer within the Hydrogen Bonding Network of Water Following Resonant Terahertz Excitation*, Sci. Adv. **6**, (2020).
[16]    A. Stupakiewicz, C. S. Davies, K. Szerenos, D. Afanasiev, K. S. Rabinovich, A. V. Boris, A. Caviglia, A. V.

Kimel, and A. Kirilyuk, *Ultrafast Phononic Switching of Magnetization*, Nat. Phys. **17**, 489 (2021).

[17] D. Afanasiev, J. R. Hortensius, B. A. Ivanov, A. Sasani, E. Bousquet, Y. M. Blanter, R. V. Mikhaylovskiy, A. V. Kimel, and A. D. Caviglia, *Ultrafast Control of Magnetic Interactions via Light-Driven Phonons*, Nat. Mater. **20**, 607 (2021).

[18] A. S. Disa, J. Curtis, M. Fechner, A. Liu, A. von Hoegen, M. Först, T. F. Nova, P. Narang, A. Maljuk, A. V. Boris, B. Keimer, and A. Cavalleri, *Photo-Induced High-Temperature Ferromagnetism in YTiO3*, Nature **617**, 73 (2023).

[19] A. Leitenstorfer, S. Hunsche, J. Shah, M. C. Nuss, and W. H. Knox, *Detectors and Sources for Ultrabroadband Electro-Optic Sampling: Experiment and Theory*, Appl. Phys. Lett. **74**, 1516 (1999).

[20] H. Zhao, Y. Tan, T. Wu, G. Steinfeld, Y. Zhang, C. Zhang, L. Zhang, and M. Shalaby, *Efficient Broadband Terahertz Generation from Organic Crystal BNA Using near Infrared Pump*, Appl. Phys. Lett. **114**, 241101 (2019).

[21] T. Seifert, S. Jaiswal, U. Martens, J. Hannegan, L. Braun, P. Maldonado, F. Freimuth, A. Kronenberg, J. Henrizi, I. Radu, E. Beaurepaire, Y. Mokrousov, P. M. Oppeneer, M. Jourdan, G. Jakob, D. Turchinovich, L. M. Hayden, M. Wolf, M. Münzenberg, M. Kläui, and T. Kampfrath, *Efficient Metallic Spintronic Emitters of Ultrabroadband Terahertz Radiation*, Nat. Photonics **10**, 483 (2016).

[22] C. Vicario, C. Ruchert, F. Ardana-Lamas, P. M. Derlet, B. Tudu, J. Luning, and C. P. Hauri, *Off-Resonant Magnetization Dynamics Phase-Locked to an Intense Phase-Stable Terahertz Transient*, Nat. Photonics **7**, 720 (2013).

[23] Y. Mukai, H. Hirori, T. Yamamoto, H. Kageyama, and K. Tanaka, *Nonlinear Magnetization Dynamics of Antiferromagnetic Spin Resonance Induced by Intense Terahertz Magnetic Field*, New J. Phys. **18**, 013045 (2016).

[24] S. Bonetti, M. C. Hoffmann, M. J. Sher, Z. Chen, S. H. Yang, M. G. Samant, S. S. P. Parkin, and H. A. Dürr, *THz-Driven Ultrafast Spin-Lattice Scattering in Amorphous Metallic Ferromagnets*, Phys. Rev. Lett. **117**, 087205 (2016).

[25] M. Shalaby, A. Donges, K. Carva, R. Allenspach, P. M. Oppeneer, U. Nowak, and C. P. Hauri, *Coherent and Incoherent Ultrafast Magnetization Dynamics in 3d Ferromagnets Driven by Extreme Terahertz Fields*, Phys. Rev. B **98**, 014405 (2018).

[26] B. C. Choi, J. Rudge, K. Jordan, and T. Genet, *Terahertz Excitation of Spin Dynamics in Ferromagnetic Thin Films Incorporated in Metallic Spintronic-THz-Emitter*, Appl. Phys. Lett. **116**, 132406 (2020).

[27] T. G. H. Blank, K. A. Grishunin, E. A. Mashkovich, M. V. Logunov, A. K. Zvezdin, and A. V. Kimel, *THz-Scale Field-Induced Spin Dynamics in Ferrimagnetic Iron Garnets*, Phys. Rev. Lett. **127**, 037203 (2021).

[28] R. Salikhov, I. Ilyakov, L. Körber, A. Kákay, R. A. Gallardo, A. Ponomaryov, J. C. Deinert, T. V. A. G. de Oliveira, K. Lenz, J. Fassbender, S. Bonetti, O. Hellwig, J. Lindner, and S. Kovalev, *Coupling of Terahertz Light with Nanometre-Wavelength Magnon Modes via Spin–Orbit Torque*, Nat. Phys. **19**, 529 (2023).

[29] S. Sederberg, D. Zimin, S. Keiber, F. Siegrist, M. S. Wismer, V. S. Yakovlev, I. Floss, C. Lemell, J. Burgdörfer, M. Schultze, F. Krausz, and N. Karpowicz, *Attosecond Optoelectronic Field Measurement in Solids*, Nat. Commun. **11**, 1 (2020).

[30] J. Faure, J. Van Tilborg, R. A. Kaindl, and W. P. Leemans, *Modelling Laser-Based Table-Top THz Sources: Optical Rectification, Propagation and Electro-Optic Sampling*, Opt. Quantum Electron. **36**, 681 (2004).

[31] T. Kampfrath, J. Nötzold, and M. Wolf, *Sampling of Broadband Terahertz Pulses with Thick Electro-Optic Crystals*, Appl. Phys. Lett. **90**, 231113 (2007).

[32] A. K. Zvezdin and V. A. Kotov, *Modern Magnetooptics and Magnetooptical Materials* (Taylor & Francis Group, New York, 1997).

[33] G. Neuber, R. Rauer, J. Kunze, T. Korn, C. Pels, G. Meier, U. Merkt, J. Bäckström, and M. Rübhausen, *Temperature-Dependent Spectral Generalized Magneto-Optical Ellipsometry*, Appl. Phys. Lett. **83**, 4509 (2003).

[34] P. Novák, *4.1.5 Optical and Magnetooptical Properties Landolt-Börnstein - Group III Condensed Matter 27E (Garnets)* (Springer-Verlag Berlin Heidelberg, 1991).

[35] A. Frej, C. S. Davies, A. Kirilyuk, and A. Stupakiewicz, *Laser-Induced Excitation and Decay of Coherent Optical Phonon Modes in an Iron Garnet*, J. Magn. Magn. Mater. **568**, 170416 (2023).

[36] A. L. Chekhov, Y. Behovits, J. J. F. Heitz, C. Denker, D. A. Reiss, M. Wolf, M. Weinelt, P. W. Brouwer, M. Münzenberg, and T. Kampfrath, *Ultrafast Demagnetization of Iron Induced by Optical versus Terahertz Pulses*, Phys. Rev. X **11**, 041055 (2021).

[37] Y. U. Berozashvili, S. Machavariani, A. Natsvlishvili, and A. Chirakadze, *Dispersion of the Linear Electro-Optic Coefficients and the Non-Linear Susceptibility in GaP*, J. Phys. D. Appl. Phys. **22**, 682 (1989).

[38] T. G. H. Blank, K. A. Grishunin, E. A. Mashkovich, M. V. Logunov, A. K. Zvezdin, and A. V. Kimel, *THz-Scale Field-Induced Spin Dynamics in Ferrimagnetic Iron Garnets*, Phys. Rev. Lett. **127**, 037203 (2021).

[39] W. L. Faust and C. H. Henry, *Mixing of Visible and Near-Resonance Infrared Light in GaP*, Phys. Rev. Lett. **17**, 1265 (1966).

[40] O. Gueckstock, O. Gueckstock, L. Nádvorník, L. Nádvorník, L. Nádvorník, T. S. Seifert, T. S. Seifert, M. Borchert, M. Borchert, G. Jakob, G. Schmidt, G. Woltersdorf, M. Kläui, M. Wolf, T. Kampfrath, and T. Kampfrath, *Modulating the Polarization of Broadband Terahertz Pulses from a Spintronic Emitter at Rates up to 10 KHz*, Optica **8**, 1013 (2021).

[41] G. Lezier, P. Koleják, J. F. Lampin, K. Postava, M. Vanwolleghem, and N. Tiercelin, *Fully Reversible Magnetoelectric Voltage Controlled THz Polarization Rotation in Magnetostrictive Spintronic Emitters on PMN-PT*, Appl. Phys. Lett. **120**, (2022).

[42] K. Garello, C. O. Avci, I. M. Miron, M. Baumgartner, A. Ghosh, S. Auffret, O. Boulle, G. Gaudin, and P. Gambardella, *Ultrafast Magnetization Switching by Spin-Orbit Torques*, Appl. Phys. Lett. **105**, (2014).

[43] L. Brandt, U. Ritzmann, N. Liebing, M. Ribow, I. Razdolski, P. Brouwer, A. Melnikov, and G. Woltersdorf, *Effective Exchange Interaction for Terahertz Spin Waves in Iron Layers*, Phys. Rev. B **104**, 094415 (2021).

[44] K. Neeraj, N. Awari, S. Kovalev, D. Polley, N. Zhou Hagström, S. S. P. K. Arekapudi, A. Semisalova, K. Lenz, B. Green, J. C. Deinert, I. Ilyakov, M. Chen, M. Bawatna, V. Scalera, M. D'Aquino, C. Serpico, O. Hellwig, J. E. Wegrowe, M. Gensch, and S. Bonetti, *Inertial Spin Dynamics in Ferromagnets*, Nat. Phys. **17**, 245 (2020).

[45] D. L. Mills, *Nonlinear Optics* (Springer Berlin Heidelberg, Berlin, Heidelberg, 1998).

[46] T. Kampfrath, Charge-Carrier Dynamics in Solids and Gases Observed by Time-Resolved Terahertz Spectroscopy, Freie Universität Berlin, 2006.

[47] O. Maximova, S. Lyaschenko, I. Tarasov, I. Yakovlev, Y. Mikhlin, S. Varnakov, and S. Ovchinnikov, *The Magneto-Optical Voigt Parameter from Magneto-Optical Ellipsometry Data for Multilayer Samples with Single Ferromagnetic Layer*, Phys. Solid State **63**, 1485 (2021).

[48] T. J. Parker, J. E. Ford, and W. G. Chambers, *The Optical Constants of Pure Fused Quartz in the Far-Infrared*, Infrared Phys. **18**, 215 (1978).

[49] J. Zak, E. R. Moog, C. Liu, and S. D. Bader, *Fundamental Magneto-optics*, J. Appl. Phys. **68**, 4203 (1990).

[50] C. Y. You and S. C. Shin, *Generalized Analytic Formulae for Magneto-Optical Kerr Effects*, J. Appl. Phys. **84**, 541 (1998).

[51] M. Buchmeier, R. Schreiber, D. E. Bürgler, and C. M. Schneider, *Thickness Dependence of Linear and Quadratic Magneto-Optical Kerr Effects in Ultrathin Fe(001) Films*, Phys. Rev. B **79**, 064402 (2009).

# Supplementary Material for
# Broadband spintronic detection of the absolute field strength of terahertz electromagnetic pulses

A.L. Chekhov, Y. Behovits, U. Martens, B.R. Serrano, M. Wolf, T.S. Seifert, M. Muenzenberg and T. Kampfrath

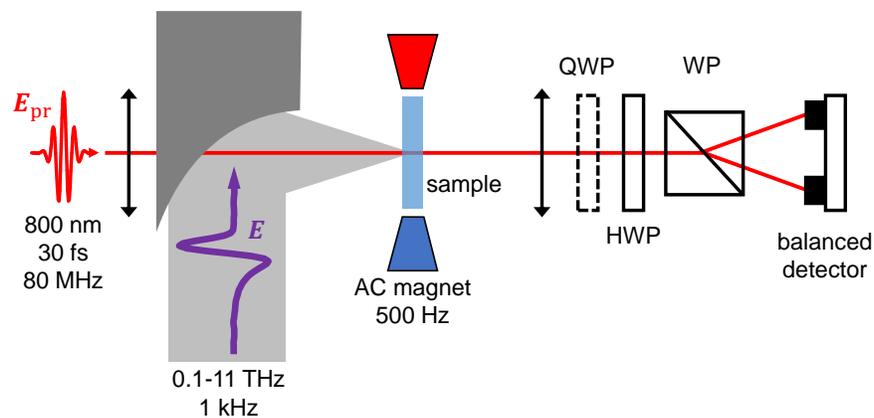

Figure S5. Schematic of the setup used for Zeeman-torque sampling (ZTS) of THz pulses. Broadband THz pulses (0.1-11 THz) are generated by excitation of a Si-based spintronic THz emitter (Si-STE, TeraSpinTec GmbH) [1] with femtosecond optical pulses from an amplified Ti:Sapphire laser system (Coherent Legend Elite Cryo PA; center wavelength 800 nm, pulse energy 5 mJ, repetition rate 1 kHz). They are detected with probe pulses from the amplifier seed oscillator (Coherent Vitara; center wavelength 800 nm, duration 30 fs). The magnetic sample is magnetized along the $x$-axis [Fig. 1(b)], which is parallel to the sample plane and plane of incidence, using a 500 Hz AC magnet. Polarization variations of the transmitted probe pulses are detected with an optional quarter-wave plate (QWP), a half-wave plate (HWP), a Wollaston prism (WP) and a balanced detector (Femto Balanced Photoreceiver).

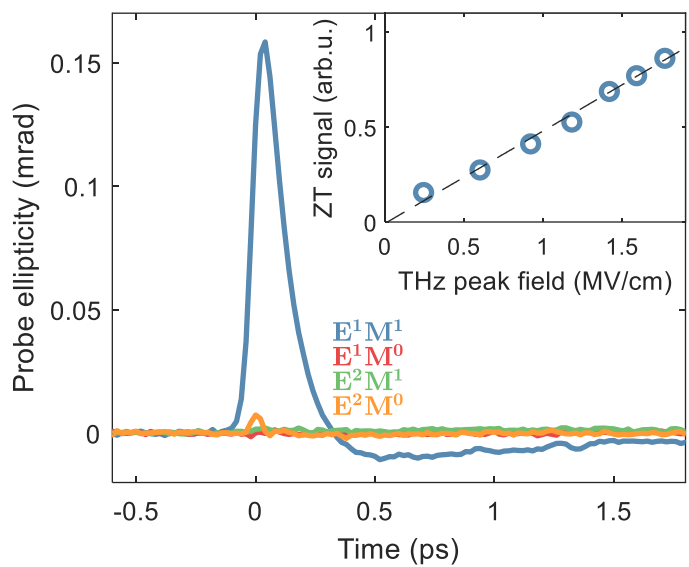

Figure S6. Various contributions to the pump-probe signal, linear or quadratic in the THz field $E$ and the sample magnetization $M_0$. The response linear in both $E$ and $M_0$ (blue solid line) dominates and is assigned to Zeeman torque, i.e., the ZTS signal. Inset: Scaling of the ZTS signal with the incident THz field strength. The dashed line is a linear fit.

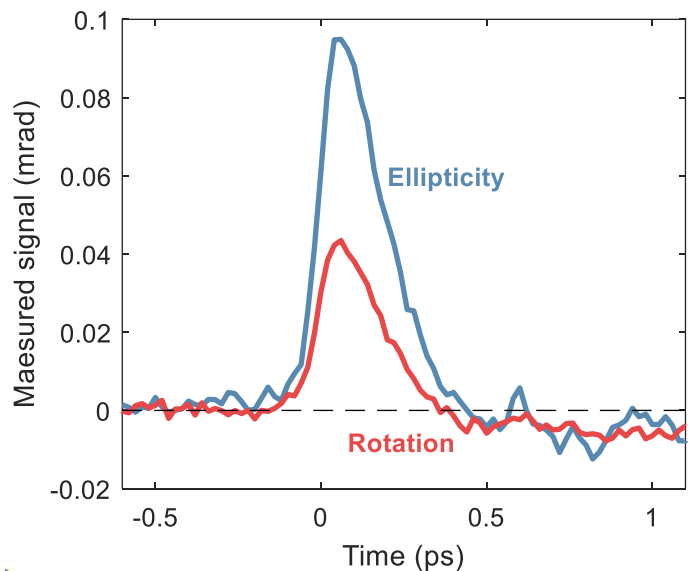

Figure S7. ZTS signals measured in the Fe(8 nm) sample for probe-polarization rotation (red) and ellipticity (blue). In this measurement, the setup was not purged, resulting in a distorted THz pulse.

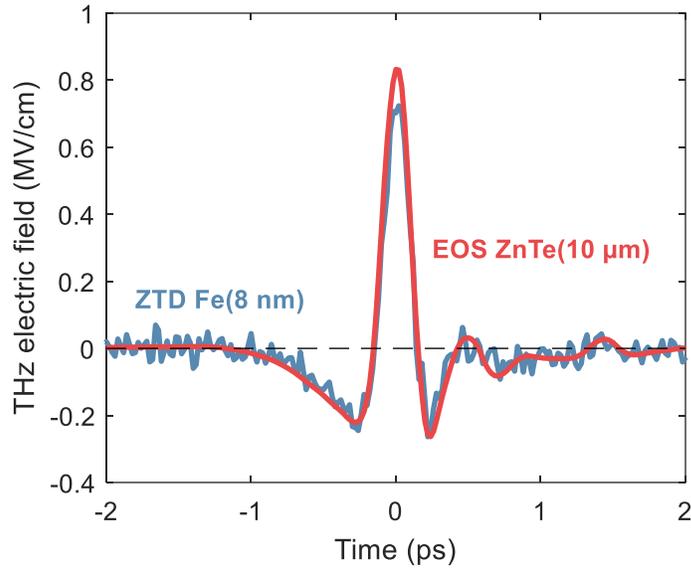

Figure S8. Absolute THz electric field of a THz pulse from the LiNbO$_3$ source of Ref. [2]. The red curve shows the transient electric field extracted from the signal of EOS in a 10 μm ZnTe(110) crystal as in Ref. [1]. The blue curve was extracted from the ZTS signal in Fe(8 nm). The sampling geometry was the same as in Fig. S1.

| Parameter | Value |
|---|---|
| $\sigma_0$ | 1.2 MS/m [2] |
| $\tau$ | 10 fs [2] |
| $n_3$ | 1.96 ( [3], 0.5-4 THz) |
| $n_2(\omega_{pr})$ | 1.4+3.8i ( [4], 12 nm Fe film) |
| $n_3(\omega_{pr})$ | 1.45 [5] |

Table S1. Values and literature sources of parameters used in the calculations

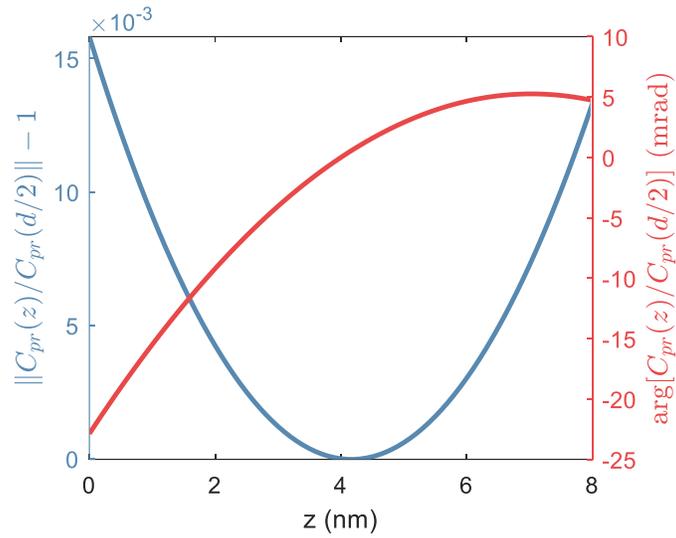

Figure S9. Amplitude (blue) and phase (red) parts of the sensitivity factor $(1 + r_{21}e^{2i\beta_2 z})(1 + r_{23}e^{2i\beta_2(d-z)})$ with respect to its value in the center ($z = d/2$) for a film with $d = 8$ nm, probe wavelength 800 nm and refractive index $n_2(\omega_{pr}) = 1.4 + 3.8i$.

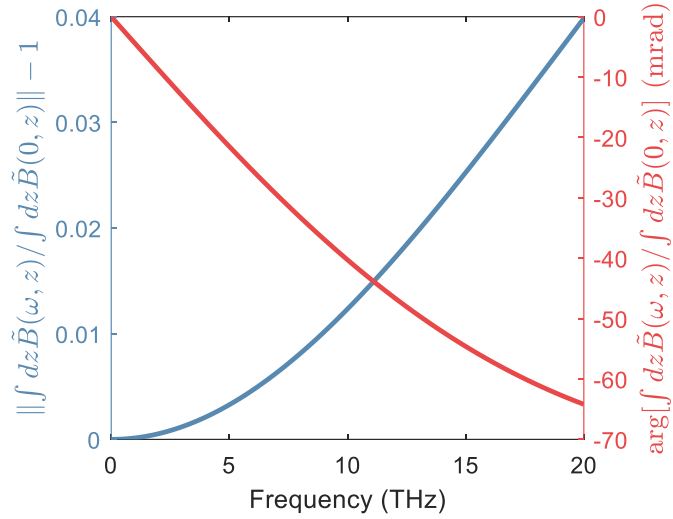

Figure S10. Frequency dependence of amplitude (blue) and phase (red) of the term in front of $\boldsymbol{v} \cdot \widetilde{\boldsymbol{E}}_{inc}$ in Eq. (10). To simulate this dependence, we use the Drude model for the THz conductivity, $\sigma(\omega) = \sigma_0/(1 - i\omega\tau)$, with mean electron scattering time $\tau$ and DC conductivity $\sigma_0$. The terahertz refractive index of glass is assumed to be constant over the studied frequency range. Amplitude and phase exhibit minor variations in the relevant range 0.1-11 THz. Table S1 shows the parameters used.

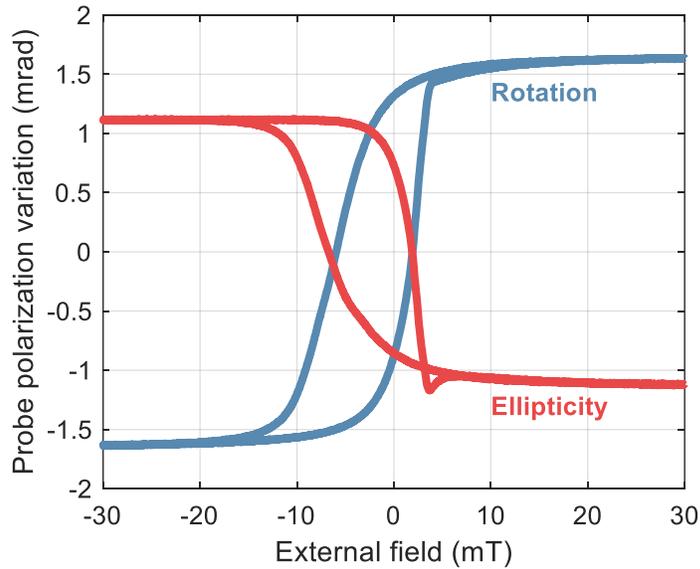

Figure S11. Static hysteresis loops for longitudinal magneto-optic Kerr effect measured for rotation (blue) and ellipticity (red) of the probe-beam polarization. Non-symmetric shape of the loops is attributed to even magneto-optic effects. This does not affect the extracted size of the magneto-optic Kerr effect, since we measure in saturation.

**References**

[1] R. Rouzegar, A. L. Chekhov, Y. Behovits, B. R. Serrano, M. A. Syskaki, C. H. Lambert, D. Engel, U. Martens, M. Münzenberg, M. Wolf, G. Jakob, M. Kläui, T. S. Seifert, and T. Kampfrath, *Broadband Spintronic Terahertz Source with Peak Electric Fields Exceeding 1.5 MV/Cm*, Phys. Rev. Appl. **19**, 034018 (2023).

[2] A. L. Chekhov, Y. Behovits, J. J. F. Heitz, C. Denker, D. A. Reiss, M. Wolf, M. Weinelt, P. W. Brouwer, M. Münzenberg, and T. Kampfrath, *Ultrafast Demagnetization of Iron Induced by Optical versus Terahertz Pulses*, Phys. Rev. X **11**, 041055 (2021).

[3] T. J. Parker, J. E. Ford, and W. G. Chambers, *The Optical Constants of Pure Fused Quartz in the Far-Infrared*, Infrared Phys. **18**, 215 (1978).

[4] O. Maximova, S. Lyaschenko, I. Tarasov, I. Yakovlev, Y. Mikhlin, S. Varnakov, and S. Ovchinnikov, *The Magneto-Optical Voigt Parameter from Magneto-Optical Ellipsometry Data for Multilayer Samples with Single Ferromagnetic Layer*, Phys. Solid State **63**, 1485 (2021).

[5] I. H. Malitson, *Interspecimen Comparison of the Refractive Index of Fused Silica*, JOSA **55**, 1205 (1965).